\documentclass[12pt]{iopart}

\usepackage[latin1]{inputenc}
\usepackage[babel]{csquotes}
\usepackage{graphicx}
\usepackage[english]{babel}
\usepackage{hyperref}
\usepackage{iopams}

\usepackage[T1]{fontenc}

\usepackage{wasysym}
\usepackage{subfigure}
\usepackage{gensymb}

\begin{document}
\title[Low-energy electromagnetic processes affecting free-falling test-mass charging]{Low-energy electromagnetic processes affecting free-falling test-mass charging for LISA and future space interferometers}
\author{Catia Grimani$^{1,}$$^2$, Andrea Cesarini$^2$, Michele Fabi$^{1,}$$^2$, Mattia Villani$^{1,}$$^2$}
\address{$^1$DISPEA, Universit\`a di Urbino Carlo Bo via Santa Chiara 27, 61029, Urbino, Italy}
\address{$^2$INFN - Sezione di Firenze via B. Rossi, 1 50019, Sesto Fiorentino, Florence, Italy}
\ead{mattia.villani@uniurb.it}

\begin{abstract}
Galactic cosmic rays and solar energetic particles charge gold-platinum, free-falling test masses (TMs) on board interferometers for the detection of gravitational waves in space. The charging process induces spurious forces on the test masses that affect the sensitivity of these instruments mainly below $10^{-3}$ Hz. Geant4 and FLUKA Monte Carlo simulations were carried out to study the TM charging process on board the LISA Pathfinder mission that remained into orbit around the Sun-Earth Lagrange point L1 between 2016 and 2017. While a good agreement was observed between simulations and measurements of the TMs net charging, the shot noise associated with charging fluctuations of both positive and negative particles resulted 3-4 times higher that predicted. The origin of this mismatch was attributed to the propagation of  electrons and photons only above 100 eV in the simulations. In this paper, low-energy electromagnetic processes to be included in the future Monte Carlo simulations for LISA and LISA-like space interferometers TM charging are considered. {It is found that electrons and photons below 100 eV give a contribution to the effective charging comparable to that of the whole sample of particles above this energy. In particular, for incident protons ionization contributes twice with respect to low energy kinetic emission and electron backscattering. The other processes are found to play a negligible role. For heavy nuclei only sputtering must be considered.}
\end{abstract}
\noindent{\it Keywords\/}: Gravitational radiation detectors; Cosmic rays; Electromagnetic processes and properties

\submitto{\CQG}
\maketitle

\section{Introduction}

LISA Pathfinder (LPF) \cite{lpf1,lpf2,lpf3,lpf4} was the European Space Agency mission aiming to test  the technology that will be placed on board LISA \cite{lisa} (Laser Interferometer Space Antenna), the first  space-based  gravitational wave observatory scheduled to be sent to space in 2034. The LPF spacecraft (S/C) was launched on December 3, 2015 from the Kourou base in French Guiana and reached its final orbit around the L1 Lagrangian point of the Earth-Sun system at 1.5 million kilometers from the Earth by the end of January 2016. {The mission ended July 18, 2017. }

Two cubic  gold-platinum test masses (TMs) of nearly 2 kg mass each and 4.6 cm side constituted the heart of the LPF inertial sensor  playing the role of mirrors of the on-board interferometer. A gold coating of 0.9 $\mu$m was deposited on the TMs and electrodes. A high density, low-magnetic susceptivity  alloy was chosen for the TMs. These proof masses remained  nominally in free-fall during the mission, controlled by a gravity reference system (GRS). The GRS was constituted by electrodes surrounding the TMs at a distance of about 3 mm; the whole system was enclosed in a vacuum  chamber, the \emph{vacuum enclosure}, vented to the outside. Sensing electrodes allowed to measure the position of the TMs through the variation of the system capacity, while control electrodes actuated separately each TM in order to keep it to selected positions within the electrode-housing. 

Spurious coulombian forces act on the TMs due to the charging process of galactic cosmic rays (GCRs) and solar energetic particles (SEPs) traversing or interacting in  13.8 g cm$^{-2}$ of S/C material. This process was estimated to constitute one of the most relevant sources of noise below $10^{-3}$ Hz mainly during SEP events for LISA-like interferometers \cite{vocca2004,vocca}. A system of ultraviolet lamps was placed inside the electrode housing of LPF to illuminate and discharge the TMs with $\approx 4.88$ eV energy photons via the photoelectric effect \cite{UV1,UV2}. Unfortunately, TM charging and discharging during SEP events was not tested on board LPF since no SEPs overcoming the GCR background were observed during the mission elapsed time. A test-mass discharging system is also under development for LISA \cite{IC2}. LPF proved that the technology for gravitational wave space interferometers is mature. As a matter of fact, the residual acceleration noise on the TMs was found of the order of femto-g, approximately one order of magnitude lower than required \cite{lpf4} with a GCR flux bombardment near maximum as observed during the declining phase of the solar cycle 24.

In addition to the average predictions of the GCR intensity associated with the solar activity of the solar cycle 26 \cite{sol_cyc}, an accurate radiation environmental study for the TM charging will be carried out for LISA involving  recurrent and non recurrent GCR short-term variations ($<$ 1 month)  generating extra noise force on the TMs. Before the LPF mission launch, the TM charging was studied with Monte Carlo simulations \cite{IC,wass,nostro2,nostro}. The net charging prediction associated with GCR only, resulted in  good agreement with LPF measurements, while the shot noise was  under-estimated by a factor 3-4 \cite{measured,mio}. In order to investigate the origin of this mismatch,  the GCR spectra adopted in the simulations were first compared to the international space station AMS-02 magnetic spectrometer experiment \cite{AMS} measurements carried out at the time the mission was in orbit. An agreement was found within $\pm$10 \% \cite{mio}. According to preliminary results \cite{mio} the lack of propagation of low-energy electrons in both FLUKA  ($<1$ keV) \cite{fluka,fluka2} and Geant4 Monte Carlo toolkit  ($<100 $ eV) \cite{geant} may explain the difference between measured and observed charging noise. {In particular, despite eV electron range in gold is of the order of nanometers, particles produced at the separation region between TMs and electrode housing may play an important role.}

The aim of this paper is to discuss the main electromagnetic processes leading to the production of low-energy electrons. In particular, ionization, kinetic emission of electrons, atomic sputtering and photoelectric effect associated with low-energy  photons are discussed. Quantum mechanics (QM) processes limiting the number of electrons charging the TMs are evaluated.  Those processes found to give a non-negligible contribution to the LISA TM charging will be added to Geant4 and FLUKA Monte Carlo simulations. The same is recommended for the future LISA-like space interferometers \cite{decigo,wtni}.

For a previous paper on these topics,  see \cite{IC}. However, in that work the contribution of $Z>2$ GCR nuclei, found here to be crucial, was not considered.

This paper is organized as follows: in Section \ref{sec:ionization} the production of low-energy electrons by ionization energy loss of cosmic-ray ions traversing the S/C is discussed. In section \ref{sec:kinetic} the kinetic emission of electrons due to the bombardment of fast but not relativistic particles (with energies of the order of keV-MeV) is presented; in section \ref{sec:sputtering} the atomic sputtering is evaluated; in Section \ref{sec:fotoni} the photoelectron production of low-energy photons is illustrated; in Section \ref{sec:QMeff} QM effects such as low-energy electron diffraction and backscattering are described. {In Section \ref{sec:summary} the contributions of each process considered in this work to the LISA TM effective charging are summarized.} finally a discussion is reported in section \ref{sec:conc}.

\section{LISA Pathfinder pre-launch test mass charging simulations and in-orbit measurements}
\label{sec:effective}

The LPF TM net ($\lambda_{net}$) and effective charging ($\lambda_{eff}$) are defined as follows (see \cite{nostro}):
\begin{equation}
\lambda_{net} = \sum_{j_{net}=-\infty}^{+\infty} j_{net}\lambda_{j_{net}}  \qquad \lambda_{eff} = \sum_{j_{eff}=-\infty}^{+\infty} (j_{eff})^2\lambda_{j_{eff}} 
\end{equation}
where $j_{net}$ is the net charge deposited  by each cosmic-ray particle interaction calculated by algebraically summing up positive and negative charges and $j_{eff}$ is the effective charge deposited in terms of equivalent single charges by considering both positive and negative charges. $\lambda_{j_{net}}$ and $\lambda_{j_{eff}}$ are the occurrence rate of the associated deposited charges {(expressed in e s$^{-1}$)}. The shot noise is calculated with the effective charging as follows \cite{nostro}:
\begin{equation}
S=\sqrt{2e\lambda_{eff}} \quad e \, s^{-1} \, Hz^{-1/2}.
\end{equation}

{The net and effective charging measured with LPF \cite{measured} are reported in table \ref{tab:mis_sim} along with the simulations results carried out before the mission launch \cite{nostro}}. The range of predictions in the simulations was associated with minimum and maximum solar modulation and corresponding GCR intensities estimated for the first part of the LPF mission. The solar activity, modulates the GCR flux observed in the inner heliosphere below 10 GeV \cite{merida}. The solar modulation observed in spring 2016 corresponded to the average between minimum and maximum predictions. As a result, we were expecting to observe a net charging of about +23 e s$^{-1}$ where $e=1.602\times 10^{-19}$ C and an effective charging of approximately 232 e s$^{-1}$. It is possible to notice that the net charging was in good agreement with the net charge measurements, while the effective charging was severely underestimated.

A wrong estimate of the incident GCR fluxes would have led to a mismatch similar for both net and effective charging, moreover after GCR data for the beginning of 2016 \cite{AMS} were published, we found that our predictions were in agreement within 10\% with observations. The possibility that the effective charging mismatch was due to the lack of propagation of electrons below 100 eV was explored in \cite{mio} and appeared very promising. In particular, low-energy protons and nuclei traversing the electrodes are observed to produce a large number of electrons and both nuclei and electrons are observed to stop in the TM giving a large contribution to the effective charging.

\begin{table}[ht]
\centering
\begin{tabular}{c|c|c}
& Net charge rate (e s$^{-1}$) & Effective charge rate (e s$^{-1}$)\\
\hline
Predicted & $15.3-38.2$ & $171.3-311.8$\\
\hline
Measured & TM1 22.1 $\pm$ 1.7 & TM1 1060 $\pm$ 90\\
& TM2 23.4 $\pm$ 2.1 &TM2 1360 $\pm$ 130
\end{tabular}
\caption{LPF TM charging simulations \cite{nostro} and measurements \cite{measured}. In \cite{nostro}, simulations were carried out several months before the mission launch with the solar modulation conditions expected for the mission. The measurements for both the TMs reported in \cite{measured} were carried out on April 20-23, 2016}\label{tab:mis_sim}
\end{table}

\section{Ionization low-energy electron production} 
\label{sec:ionization}

In the Monte Carlo simulations carried out with Geant4 \cite{geant} and FLUKA \cite{fluka,fluka2} for LPF, the generation of electrons was {allowed down} to the mean ionization potential of the absorbing material (790 eV in gold) and to 1 keV, respectively. This was done in order to reduce the CPU time required for the simulations. In general, this is considered a good compromise, however this was not the case for LPF, because of the high sensitivity of the interferometer. {In \cite{vocca2004,vocca} the LISA  TM charging was discussed by considering both incident GCRs and SEPs. Despite a simplified geometry of the S/C was considered and no low-energy propagation was included in the Monte Carlo, it was shown that an increasing flux of incident solar particles was limiting the mission sensitivity below 1 mHz. The considered higher incidence flux of low energy electrons would play a similar role when GCR particles are simulated and would sum up to the SEP contribution.} 

An ionizing particle traversing a metal produces free electrons whose number can be calculated as a function of the charge and of the energy of the incident particle. The majority of ionization energy loss parameterizations are available for incident particles energy above 10 keV (see for example \cite{leo,pdb}). However, in \cite{cuci,cuci2,cuci3} a semi-empirical parametrization of the energy lost through ionization by protons ions and electrons valid also at lower energies is presented:
\begin{equation}\label{eq:cucinotta}
\frac{dn}{dK}= \frac{2\pi N e^4}{mc^2 \beta^2} \frac{Z_s^2}{K^2} \left[ 1- \frac{\beta^2E}{E_m} + \frac{\pi \beta Z_s^2}{137} \sqrt{\frac{K}{E_m}} \left( 1- \frac{K}{E_m} \right) \right] dx
\end{equation}
where $K$ is the kinetic energy of the electron, $\beta=v/c$ is the particle velocity, $N$ is the electron density of the material and $dx$ its thickness;
\begin{equation}\label{eq:Emax}
E_m=\frac{2mc^2\beta}{1-\beta^2}
\end{equation}
is the maximum energy transferred to an electron, $e$ and $m$ are the electron charge and mass, respectively and, finally, the effective charge of a particle of atomic number $Z$ is given by:
\begin{equation}
Z_s=Z\left[ 1- \exp\left( -\frac{125\beta}{Z^{2/3}} \right) \right].
\end{equation}
This parameterization was used in \cite{cuci,cuci2,cuci3}  to calculate the energy deposited by ionizing particles traversing different materials like dielectrics and metals as a function of the material thickness. Observations were compared to calculations and  a good agreement with  experimental data down to 10 eV was found for the aforementioned particles. {For pions the ionization energy loss reported in \cite{pion1,pion2,pion3} was adopted.}

Millions of low-energy electrons are produced by ionization. {However, electrons with energies of a few electronvolts typically propagate through matter for a few nanometers} being quickly reabsorbed by the material. In the LPF case, only those electrons generated very close to the surface can leave the TMs or the electrodes, thus contributing to the net and effective charging. A preliminary study was carried out in \cite{mio}. The major contribution to the TM charging is given by protons and helium nuclei which constitute about 98\% of the GCR sample. Although nuclei represent only 1\% of the GCRs, their contribution to the production of low-energy electrons is also significant because of their high electric charge, as results from  equation \eref{eq:cucinotta}. As an example, a 100 MeV proton and a nucleus of iron of 100 MeV/n escaping the electrodes surrounding the TMs and stopping in the TMs would generate on average,  0.1 and 820 electrons, respectively. Despite these examples focus only on stopping ionizing particles and hadronic interaction in the S/C would mitigate their impact on the overall charging process, ionization appears to be the most relevant process to explain the mismatch between LPF measured and simulated shot noise.

In order to estimate the effects of low-energy electrons and other secondary particles on the TM charging process, preliminary simulations were carried out with FLUKA by considering a S/C simplified geometry after the LPF mission end \cite{mio}. Protons, helium  nuclei, carbon nuclei and iron nuclei were considered with energies between 100 MeV and 100 GeV impinging on a slab of 13.8 g cm$^{-2}$ of aluminum at different angles ranging from 0$^\circ$ to 80$^\circ$. Carbon was chosen as representative of the CNO group (the most abundant of the GCR ion bulk) and iron because of its high charge. The resulting secondary electrons  were propagated in the LPF electrode housing with the Monte Carlo  LEI (Low Energy Ionization, first presented in  \cite{mio}) that allows for the production and propagation of low-energy electrons down to 12 eV (see section \ref{sec:QMeff} for details about this energy cut-off). As a result of these simulations, it was found that protons incident on the S/C with energies below 100 MeV are stopped by the material before reaching the TMs, while for heavier nuclei the stopping energies increased to 600 MeV/n. Protons and nuclei with energies well above 100 MeV at the top of S/C produce a large number of secondaries, mostly pions and other mesons such as K and $\eta$ and muons. {However, electrons and positrons represent the main components of charged particles affecting the TM effective charging. Moreover, high-Z nuclei, iron nuclei for instance}, at energies above 1 GeV/n produce a number of secondaries of about one order of magnitude larger than the number of incoming nuclei. Pions produced by protons have been found to increase by 14\% the charge deposited on the TM by protons alone. Given the encouraging results obtained with this preliminary work, new simulations are in preparation for LISA were the FLUKA Monte Carlo outcomes will be used as input data for the LEI program. Simplified geometries will be considered until the LISA S/C arrangement and materials will not be available. {New Geant4-based simulations will be also carried out in the near future.}

As pointed out above, ionization of low-energy electrons are supposed to give a contribution to solve the mismatch between former Monte Carlo simulations and LPF measurements of the TM charging. However, in the following other low-energy electromagnetic processes are considered for completeness.

\section{Kinetic electron emission}
\label{sec:kinetic}

Low-energy electrons may escape from the TMs due to bombardment of non relativistic incident particles (with energies of the order of keV-MeV). According to incident particles species this
process is named IIEE, from \emph{ion induced electron emission} or EIEE, from \emph{electron induced electron emission}.

The kinetic electron emission is a three step process: first, an internal electron is excited by an incoming low-energy particle, then the electron propagates into the medium losing energy by ionization and finally reaches the surface of the material; if its energy exceeds the potential barrier present at the surface it can escape into the external void. Several quantum processes can excite an internal electron, such as scattering, both elastic and inelastic, decay of plasmons, excitation of core electrons and Auger process  (see \cite{libro_see,schou}); it was demonstrated in \cite{cesarini} that the excitation of surface phonon does not contribute to the electron emission. The electron yield (indicated by $\delta$ for EIEE and by $\gamma$ for IIEE), defined as the total number of electrons divided by the total number of incident particles, sums up the result of all these processes.

Following the work by Schou \cite{schou} (see also \cite{IC,libro_see}),  in the case the beam of primary electrons is perpendicular to the surface of the material, the yield of the EIEE is given by:
\begin{equation}\label{eq:y_el}
\delta(E)=\beta^*S_e(E) \Lambda(1+\kappa \eta(E))
\end{equation}
where $\beta^*$ is a function that depends on the incoming particle and material, but is almost independent of the energy; $S_e(E)$ is the electron stopping power at the energy $E$ (for electron energy lower than 10 keV can be found in \cite{SPE}, while above 10 keV can be calculated with the Bethe-Bloch formula corrected by radiative processes such as bremsstrahlung \cite{leo}); $\Lambda$ is a function that depends only on the material; the parameter $\kappa$ is the backscattering efficiency and the function $\eta(E)$ is the amount of  secondaries generated by incident electrons \cite{IC,libro_see,eta}. From the  Schou's paper \cite{schou} $\Lambda=0.33$ eV$^{-1}$ {\AA} for gold and the adimensional function $\beta^*$ is $1$ in the range of energies of interest of this work. $\kappa$ is equal to one \cite{IC}, while $\eta(E)$ must be parameterized as reported in \cite{eta}.

\begin{figure}[h]
\centering
\includegraphics[scale=0.5]{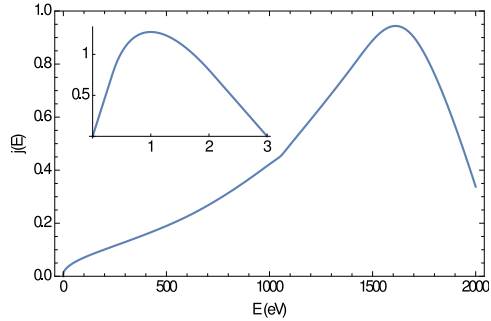}
\caption{Spectrum, $j(E)$, of electrons emitted through kinetic process produced by a beam of 2 keV electrons. The main plot shows backscattered primaries, while  the inset graph reports the spectrum of the electrons emitted through EIEE. See also \cite{IC,libro_see}.}\label{fig:je}
\end{figure}

In figure  \ref{fig:je} (see also figure 15 in \cite{IC} and figure 2.2 in \cite{libro_see}), the energy spectrum of secondary electrons emitted by 2 keV electrons perpendicularly incident on a  gold slab is shown. The curve at keV energies reported in the main graph represents elastically backscattered primaries, while particles in the eV range in the inner graph represent secondary electrons emitted from a material layer close to the surface after losing an energy equal to the work function; in the case of metals the thickness of this layer is typically 5-20 {\AA} \cite{schou}. Geant4 and FLUKA allow for the simulation of backscattered primaries down to 100 eV; future studies dedicated to the LISA TM charging will include the contribution of eVs electrons.

The case of an electron beam impinging at an arbitrary angle $\theta$ on a surface is discussed in \cite{cesarini2,cesarini3}, where an empirical parameterization of the electron yield is reported:
\begin{equation}\label{eq:angolo}
\delta(E,\theta)=\delta_{max}(\theta)\left[ B(E,\theta)\,\exp\left( 1-B(E,\theta) \right) \right]^k
\end{equation}
where:
\begin{equation}
k=\left\{ \begin{array}{cc}
k_1 & B(E,\theta)<1 \\
k_2 & B(E,\theta)>1
\end{array} \right.
\end{equation}
\begin{equation}
E_{max}(\theta)=E_{max}\left( 1+\frac{1}{\pi}\theta^2 \right)
\end{equation}
\begin{equation}
B(E,\theta)=\frac{E}{E_{max}(\theta)}
\end{equation}
\begin{equation}
\delta_{max}(\theta)=\delta_{max}\left( 1+\frac{1}{2\pi} \theta^2 \right)
\end{equation}
$k_{1,2}$, $E_{max}$ and $\delta_{max}$ are four free parameters that must be determined by fitting the electron yield of equation \eref{eq:y_el} with the expression \eref{eq:angolo} evaluated at $\theta=0$, corresponding to a perpendicular beam of electrons. For gold the parameter values are: $k_1=9.2\pm 0.9$, $k_2=0.0082 \pm 0.0001$, $E_{max}=48\pm 1$ eV and $\delta_{max}=2.443 \pm 0.004$.

In the case of ions impinging perpendicularly on a slab of gold, the electron yield for IIEE can be calculated as follows \cite{IC,libro_see,schou}:
\begin{equation}\label{eq:y_ion}
\gamma=\beta^*S_i(E) \Lambda
\end{equation}
where $S_i(E)$ is the stopping power of the ions which is found in \cite{SPI} for energies up to 12 MeV/amu, while at higher energies it must be calculated with the Bethe-Bloch formula as described in \cite{leo}. Also for IIEE $\beta^*=1$ and $\Lambda=0.33$ eV$^{-1}$ {\AA}. 

\begin{figure}[ht]
\centering
\subfigure{\includegraphics[scale=0.55]{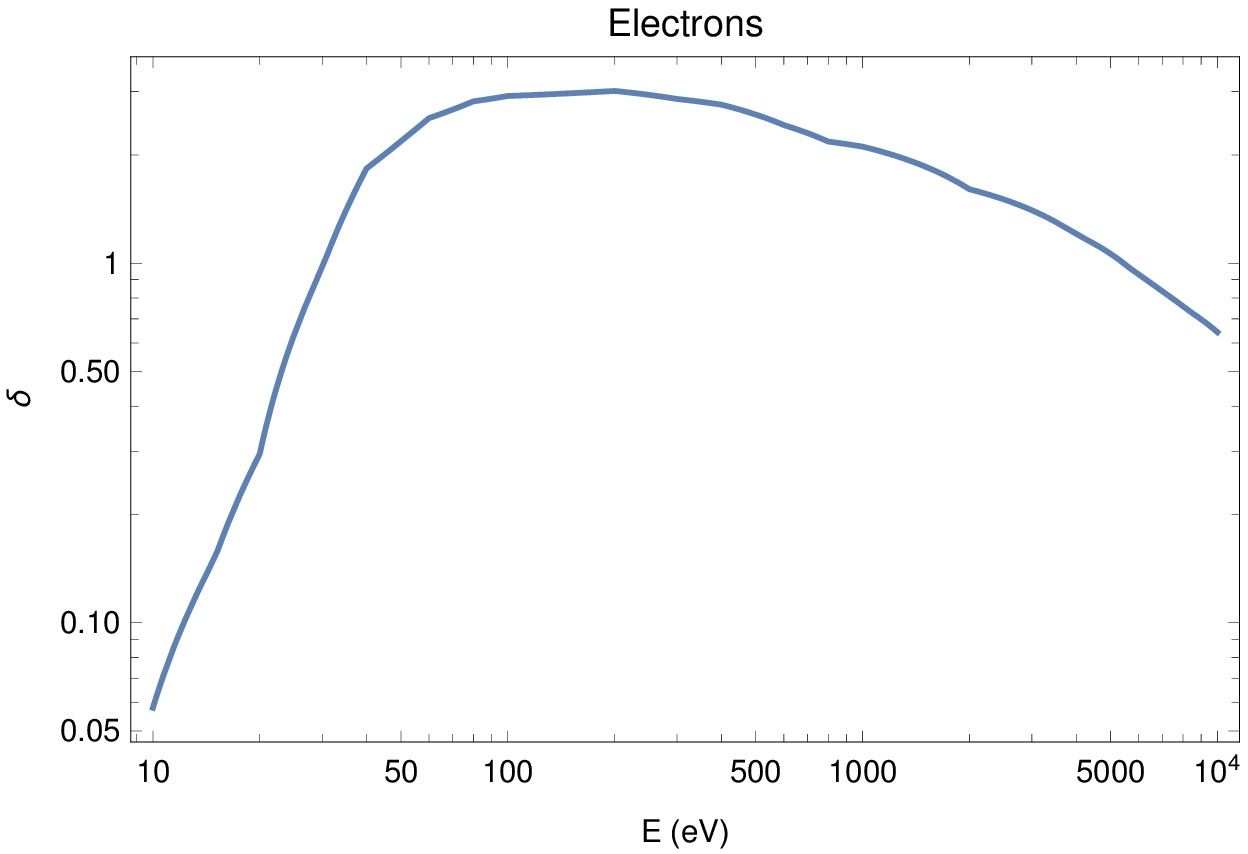}}\subfigure{\includegraphics[scale=0.75]{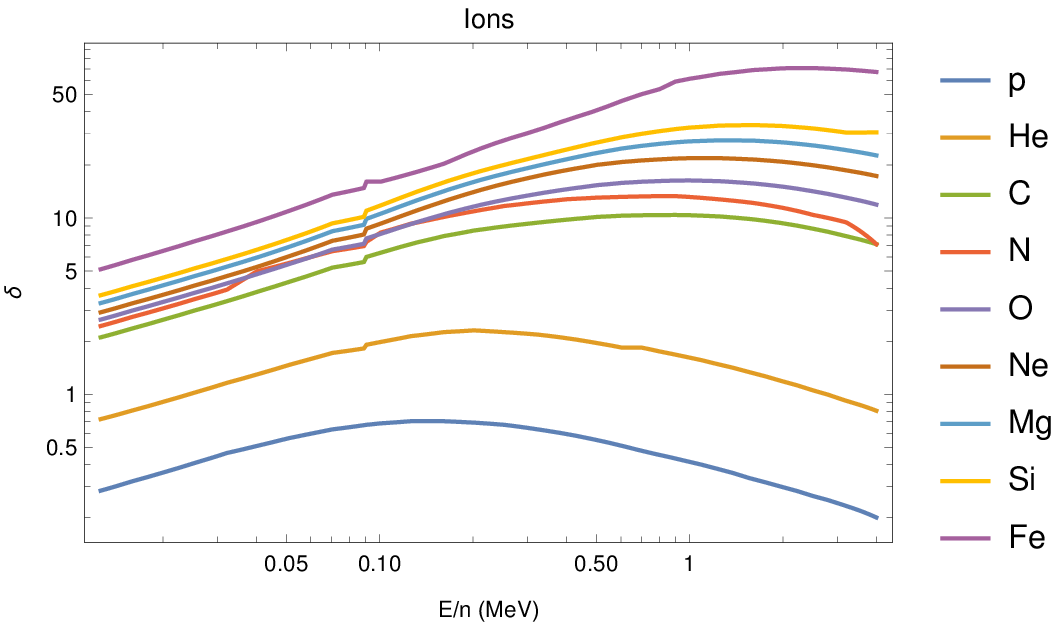}}
\caption{\emph{Left}: Electron yield of EIEE as a function of the energy of the incoming electron calculated with equations \eref{eq:y_el}. \emph{Right}: Electron yield of IIEE as a function of the energy per nucleon calculated with equation \eref{eq:y_ion} for protons, helium, carbon, nitrogen, oxygen, neon, magnesium and iron nuclei.}\label{fig:yield1}
\end{figure}

\begin{figure}
\centering
\includegraphics[scale=0.6]{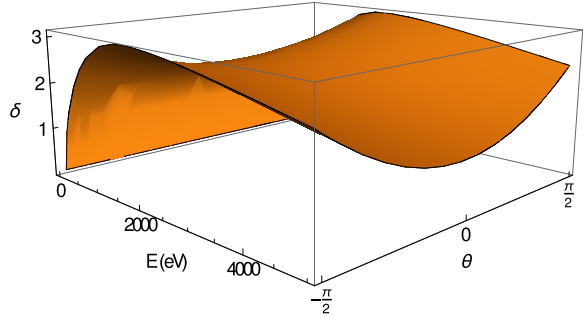}
\caption{Electron yield calculated according to equation \eref{eq:angolo} as a function of the energy and angle of incidence of the incoming electrons.}\label{fig:angolo}
\end{figure}

\begin{figure}[h]
\centering
\subfigure{\includegraphics[scale=0.7]{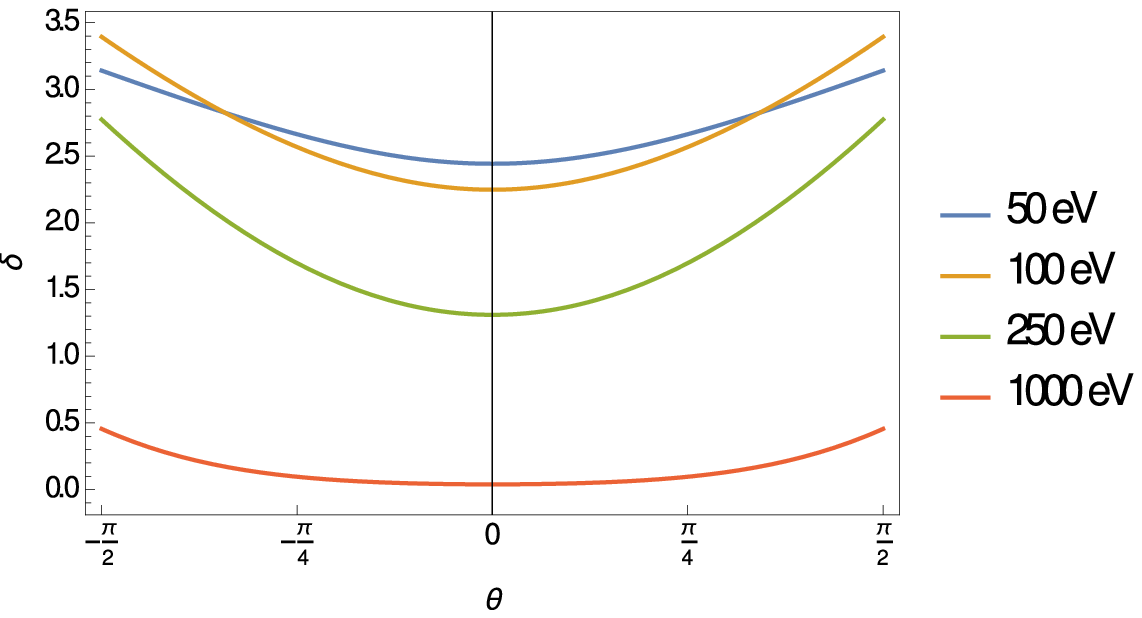}}\subfigure{\includegraphics[scale=0.7]{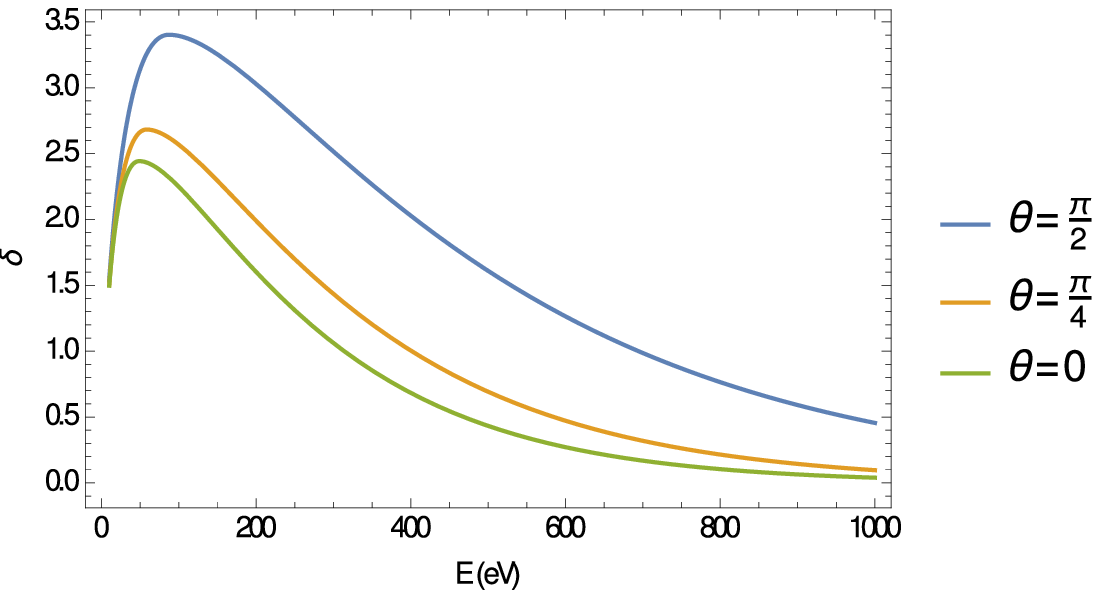}}
\caption{Same as figure \ref{fig:angolo} for the indicated energies (left) and  angles (right) of incoming electrons.}\label{fig:angolo2}
\end{figure}

{In the left panel of figure \ref{fig:yield1}, the yield of the EIEE process calculated with equation \eref{eq:y_el} is shown as a function of the incoming electron kinetic energy; in the right panel of the same figure, the yield of the IIEE process calculated with equation \eref{eq:y_ion} is reported as a function of the energy per nucleon of the incoming GCRs ions}. In figures \ref{fig:angolo} and \ref{fig:angolo2} the yield parameterized as indicated in equation \eref{eq:angolo}, depending on the energy and on the angle of incidence of the incoming electrons, as indicated in equation (7) is represented using the parameters given above. It can be observed that the yield has a minimum for $\theta=0$ (corresponding to electrons  with perpendicular incidence on the gold surface) and that at high incidence angles ($\theta\approx\pm\pi/2$),  the yield is at least twice larger than the yield at $\theta=0$. It can be concluded that the Schou formula given in equation \eref{eq:y_el} underestimates $\delta$ due to a flux of GCR primary  and secondary particles isotropically incident on LPF TMs producing electrons impinging on their surface with a random angle. 

By comparing the electron yield of protons and helium nuclei of figures \ref{fig:yield1} with those reported in figure 16b and 16c of \cite{IC}, it can be seen that the yield of the present  paper is systematically lower  than that of \cite{IC}; this is probably due to the different stopping power used in equation \eref{eq:y_ion}. {Since the yield appears $>1$ for electrons and for nuclei $\gg1$ this process must be considered for LISA and other space interferometers TM charging simulations by taking into account that the main contribution of kinetic emission is ascribable to low-energy electrons and in general to low-energy secondaries produced by GCR and SEP primary particles.}


\section{Atomic sputtering}
\label{sec:sputtering}
Atomic sputtering, the emission of atoms from a surface bombarded by ions, is also considered here because nucleus momentum transfer can cause neutral atoms or ionized matter to escape the material surface and these escaping particles may, in turn, eject electrons. Atomic sputtering is implemented neither in Geant4 nor in FLUKA. Atomic sputtering for LISA TM charging was also discussed in \cite{IC} were it was deemed as negligible for proton and helium nuclei. A more quantitative discussion is reported here. For nuclei with $Z>2$, in particular, different conclusions are taken.

This process is similar to the kinetic emission of electrons discussed above and the atomic yield $S$ is calculated as follows \cite{sputtering}:
\begin{equation}\label{eq:atom_yield}
S(x,E,\eta)=\frac{3}{4\pi^2} \frac{F(x,E,\eta)}{N C_0U_0} 
\end{equation}
where $F(x,E,\eta)$ is the nuclear stopping power depending on the energy of the incoming ion $E$, the nucleus penetration depth $x$ and the direction cosine $\eta$. $N$  is the density of atoms of the target material, $C_0$ is the constant \cite{sputtering}:
\begin{equation}
C_0=\frac{\pi}{2} \lambda_0 a
\end{equation}
where $\lambda_0=24$ and the screening radius $a=0.219$ {\AA} \cite{sputtering} are two  constants valid for all materials; finally, $U_0$ is the work function of gold (4.3 eV).

\begin{figure}[ht]
\centering
\includegraphics[scale=0.75]{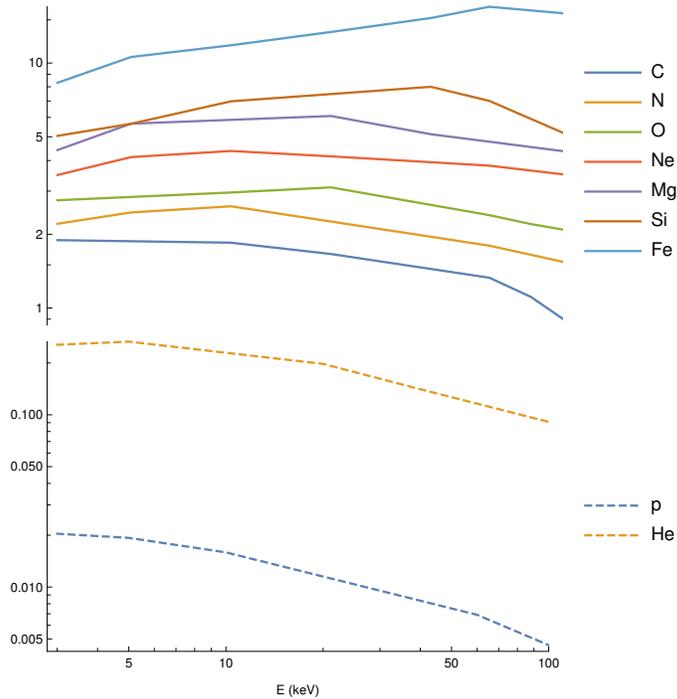}
\caption{Yield, $S$, of atom sputtered as a function of the energy per nucleon of  protons, helium, carbon, nitrogen, oxygen, neon, magnesium, silicon and iron incident on a gold slab.}\label{fig:sput1}
\end{figure}

The atom yield, $S$, has been estimated  by simulating a beam of 10000 incident ions on a 10 $\mu$m thick slab of gold with the tool SRIM \cite{srim}. Results for different GCR ions are reported in figure \ref{fig:sput1}.

The sputtering can damage the crystal lattice leading to a change of the work function and modifying the electron yield \cite{sput1,sput2,sput3,sput4}. Beams of $10^{14}-10^{15}$ ions cm$^{-2}$ of alkali metals, like lithium and cesium \cite{sput1,sput2}, or noble gases, like argon and helium \cite{sput3,sput4} were observed to change the work function of a substrate by up to 10\%; according to equation \eref{eq:y_el} this leads to a variation of +1.8\% in the electron yield if the work function decreases of 10\%, and of -1.79\% if it increases of the same amount. On the basis of the prediction of the solar cycle 26 \cite{sol_cyc} during the time the LISA mission is supposed to remain in orbit (tentatively 2034-2040),  at most a few  SEP events characterized by a fluence of $10^6-10^7$ protons cm$^{-2}$ \cite{sep}, and an overall integral particle flux of galactic origin of $10^{10}$ particles in 6 years are expected to go through the S/C. Due to the fact that ion fluxes are two orders of magnitude smaller than the proton flux it can be safely estimated that sputtering will not lead to a measurable change in the work function or electron yield. 

\begin{figure}[ht]
\centering
\includegraphics[scale=0.6]{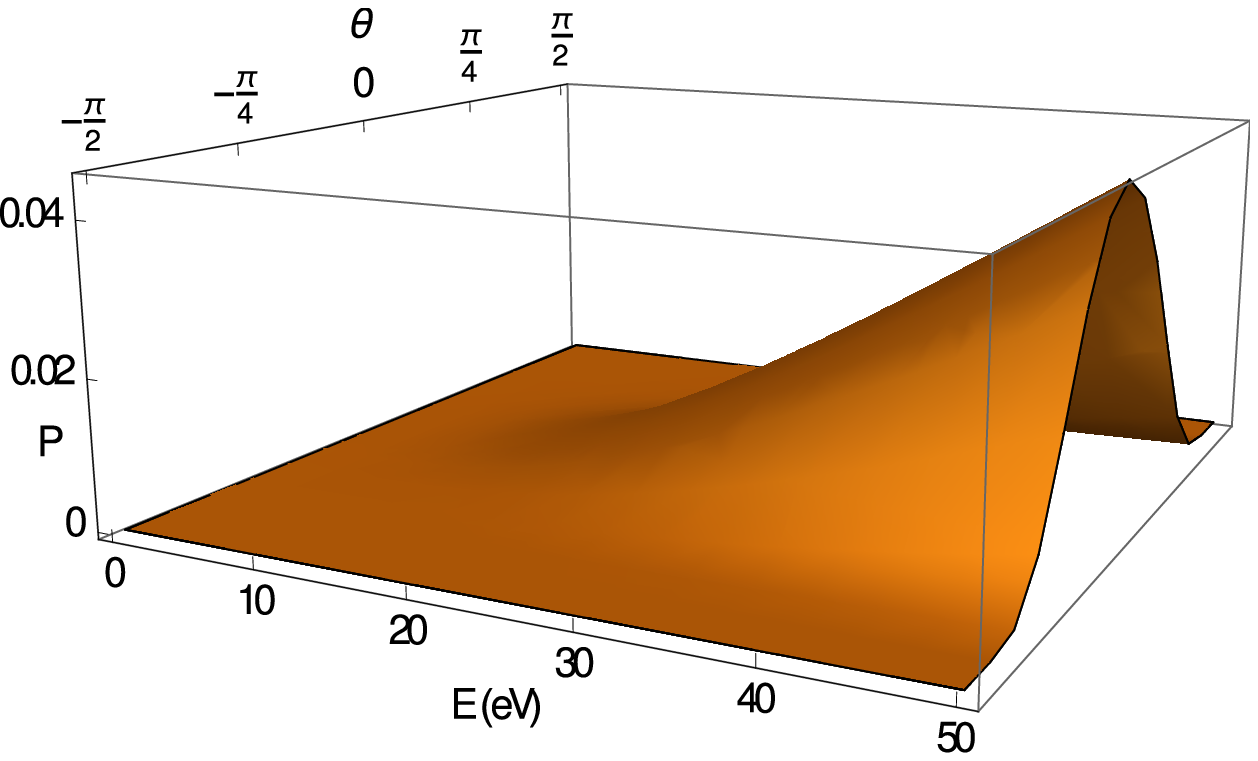}
\subfigure{\includegraphics[scale=0.7]{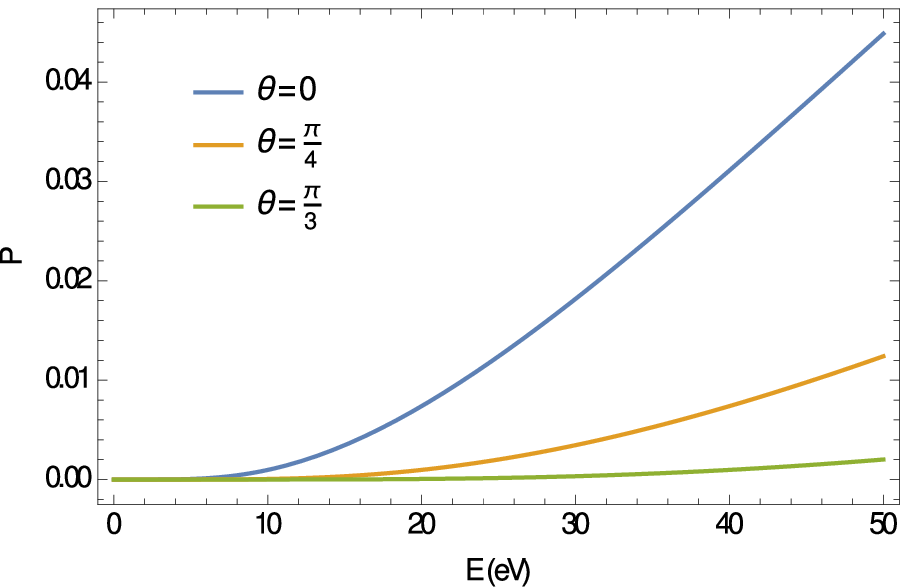}}\subfigure{\includegraphics[scale=0.7]{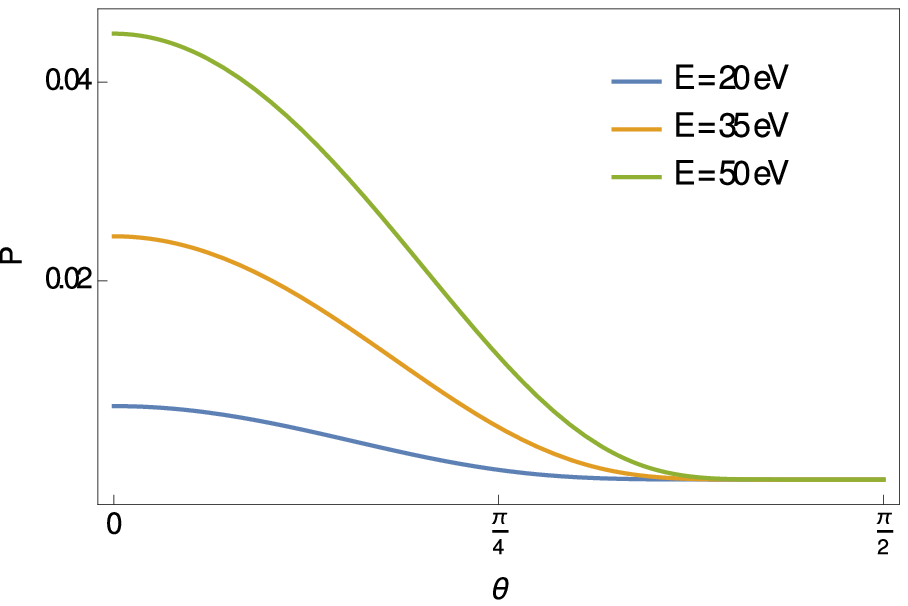}}
\caption{Top: probability of emission ($P$) of an ion due to the sputtering process as a function of the ion energy $E$ and the angle of emission $\theta$. Bottom left: same as above as a function of the energy only. Bottom right:  probability of emission of an ion as a function of the angle of emission.}\label{fig:prob}
\end{figure}

Following the work reported in \cite{sput_ion}, the probability that an ion is emitted is given by:
\begin{equation}
P=\exp\left[ -\frac{2\Delta_0}{\hbar\xi v \cos\theta} \exp\left( -\xi v t \right) \right]
\end{equation}
where $\Delta_0=2$ eV, $\xi= 1$ {\AA}$^{-1}$ \cite{sput_ion}, $v$ is the particle velocity and $t$ is time. The ion speed is estimated as indicated below:
\begin{equation}
v=c\sqrt{\frac{2K}{mc^2}}
\end{equation} 
where $c$ is the speed of light,  $K$ is the kinetic energy of the ion and mc$^2$ is the ion mass. A plot of the probability $P$ of ion emission as a function of the energy of the emitted ion and of the  angle of emission is given in figure \ref{fig:prob}: it can be observed that the probability of emission is up to 4\% for ion energies smaller than 50 eV. For nuclei this process will be considered in the future simulations of the TM charging on board LISA.

{Low-energy secondaries produced by GCR and SEP particles interacting with the TMs and surrounding electrodes will contribute to sputtering of atoms on board LISA and other space interferometers.}

\section{Low-energy photons and photoionization}
\label{sec:fotoni}

Photons are produced in several processes when energetic particles propagate through matter by neutral pions: atomic and nuclear de-excitation, bremsstrahlung, \v{C}erenkov and transition radiation. The last two  processes produce effects in transparent media only. In turn, photons generate electrons and positrons through photoelectric effect,  Compton effect and pair production. 

All of the above processes are included in Geant4 with the Livermore libraries and Penelope modules \cite{geant}, which include data from EPDL97 and EPIC2014 libraries \cite{manual}. However, the recommended low-energy limit for photon propagation is 100 eV \cite{photo,epdl97} in Geant4, while in FLUKA the propagation is inhibited for particle energies $<$ 1 keV and \emph{ad hoc} external software module  must be added.

The main processes that produce low-energy photons ($\hbar\omega<100$ eV) are \v{C}erenkov, transition radiation and bremsstrahlung; in turn  photons in this energy range produce electrons through photoelectric effect.

\subsection{\v{C}erenkov radiation}
\v{C}erenkov radiation is emitted when the velocity of a particle propagating in a medium is higher than the speed of light in that medium, which depends on the refraction index $n$ of the material. In  LPF and LISA, this radiation could be emitted in the space between the electrodes and the TM, but the gas in that region, by assuming the same conditions of LPF for LISA, has a pressure of the order of 1 $\mu$Pa and since $n\approx1$ to a good approximation, this process can be disregarded for the LISA TM charging process estimate for electrons below 100 eV.

\subsection{Transition radiation}
Transition radiation is another process leading to the production of low-energy photons. Radiation is emitted by a high-energy particle while traversing the surface of separation between two materials with different dielectric constants. The transition radiation depends on the particle energy. This process was considered also in \cite{IC}, but only in the X-ray range due to its low quantum efficiency.

In both  LPF  and LISA, the transition radiation can be produced by high-energy particles escaping the surface of electrodes and TMs, therefore the case  of a particle passing from a metal to void is considered. The  spectral intensity of the radiation is given in the following equation where $\Omega$ is the solid angle of the radiation emission and $\omega$ is the frequency of the emitted photons \cite{TR}:
\begin{equation}\label{eq:TR}
\frac{d^2I}{d\Omega d\omega}= \frac{\alpha\beta^2 Z^2}{\pi^2} \frac{\sin^2\theta \cos^2\theta}{(1-\beta^2\cos^2\theta)^2} \, \left| \frac{(\epsilon(\omega)-1)(1-\beta^2-\beta\, \sqrt{\epsilon(\omega)-\sin^2\theta})}{(\epsilon(\omega) \cos\theta+ \sqrt{\epsilon(\omega)-\sin^2\theta})(1-\beta\sqrt{\epsilon(\omega)-\sin^2\theta})} \right|^2
\end{equation}
where $Z^2$ is the atomic number of the particle, $\alpha\approx 1/137$ is the fine structure constant, $\beta=v/c$ where $v$ is the particle velocity, $\theta$ is the angle between the particle direction and the direction of the emitted photon and $\epsilon(\omega)$ is the dielectric constant of  gold. This last one can be calculated with the Drude-Sommerfeld model \cite{drude,ashcroft}:
\begin{equation}\label{eq:epsilon}
\epsilon(\omega)= 1-\frac{\omega_p^2}{\omega^2+ i  \frac{\omega}{\tau}}
\end{equation}
where $\omega_p$ is the plasma frequency of gold, $\approx 1.4\times10^{16}$ Hz, and $\tau$ is the  mean free time between two electron collisions in the metal (in the case of gold $\tau=10^{-14}$ s \cite{drude}). The number of photons emitted per second can be estimated from the double differential spectral intensity by dividing  \eref{eq:TR} for the photon energy $\hbar\omega$ (see \cite{jackson}), and by integrating over the frequency $\omega$ and the solid angle $\Omega$ in the direction of the velocity of the particle ($0\leq\theta\leq\pi/2$, $0\leq\phi\leq2\pi$), since only photons emitted in the particle direction will reach the TMs.

\begin{figure}
\centering
\includegraphics[scale=0.8]{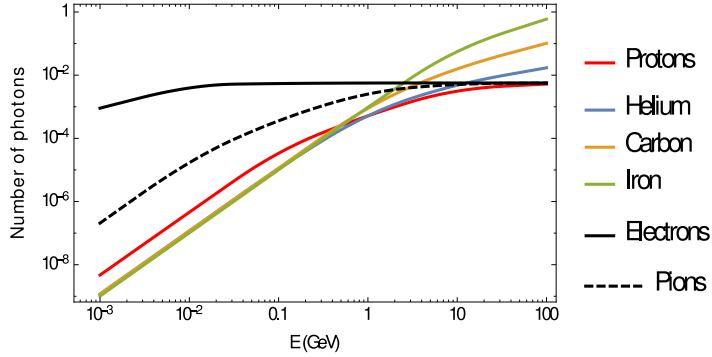}
\caption{Number of photons with energies between 4.3 eV and 100 eV emitted by protons, helium, carbon and iron  nuclei and by electrons and pions through transition radiation around the LISA TMs as a function of the total energy.}\label{fig:TR}
\end{figure}

Low-energy photons impinging on a material can cause the emission of electrons through photoelectric effect if the photon energy is larger than the working function of the material that for gold is 4.3 eV \cite{ashcroft}. The number of photons with energies ranging between 4.3 eV and 100 eV produced by a single incoming particle is reported in table \ref{tab:TR2} and in figure \ref{fig:TR} for protons, helium, carbon and iron nuclei and in table \ref{tab:TR3} for electrons and pions. {Both high-energy primary GCRs and secondary particles produced in the interaction of GCRs with the S/C material are favoured in contributing to transition radiation emission. The contribution of SEPs, characterized by lower energies than GCRs, is minor.}

\begin{table}[h!]
\centering
\begin{tabular}{c|cccccc}
Energy (GeV n$^{-1}$) & Protons & Helium & Carbon & Iron\\
\hline
0.1  & $3\times 10^{-5}$ & $10^{-4}$ & $0.001$ & $0.02$ \\ 
1    & $5\times 10^{-4}$ & $0.002$   & $0.02$  & $0.3$  \\
10   & $0.003$           & $0.01$    & $0.1$   & $2.1$  \\
100  & $0.005$           & $0.02$    & $0.2$   & $3.5$
\end{tabular}
\caption{Number of photons with energies between 4.3 eV and 100 eV emitted through transition radiation by protons, helium, carbon and iron nuclei.}\label{tab:TR2}
\end{table}

\begin{table}[h!]
\centering
\begin{tabular}{c|cccccc}
Energy (MeV) & 1 & 10 & 100 & $10^3$ & $10^4$ & $10^5$\\
\hline
Electrons    & $9\times 10^{-4}$ & 0.004             & 0.005             & 0.006 & 0.006 & 0.006  \\
Pions        & $2\times 10^{-7}$ & $2\times 10^{-5}$ & $4\times 10^{-4}$ & 0.003 & 0.005 & 0.006
\end{tabular}
\caption{Same as table \ref{tab:TR2} for electrons and pions.}\label{tab:TR3}
\end{table}

\begin{figure}
\centering
\includegraphics[scale=0.8]{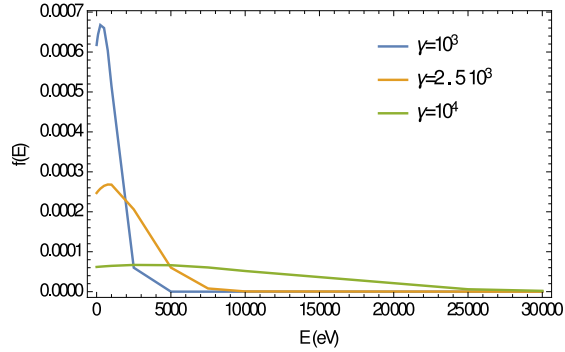}
\caption{Transition radiation photon spectrum generated by any particle with $\gamma=\{10^3,2.5\times 10^3,10^4\}$ that traverses only one surface,  calculated according to equation \eref{eq:spectrum}.}\label{fig:spectrum}
\end{figure}

The spectrum of the transition radiation emitted by a particle with $\gamma=1+\frac{K}{mc^2}$ (where $K$ is the kinetic energy of the particle) traversing \emph{only one} surface separating two different materials is given by \cite{spectrum}:
\begin{equation}\label{eq:spectrum}
f(K)=\frac{1}{\hbar \eta} \exp\left( a t-\frac{a^3}{3} \right) Ai(t)
\end{equation}
where $Ai(t)$ is the Airy function of the first kind, 
\begin{equation}
t=\frac{K-\hbar\langle\omega\rangle}{\hbar\eta}+a^2
\end{equation}
where
\begin{equation}
\eta=\sqrt[3]{\frac{\langle\omega^3\rangle}{2}}
\end{equation}
\begin{equation}
a=\eta \frac{\langle\omega^2\rangle}{\langle\omega^3\rangle}
\end{equation}
and:
\begin{equation}
\langle\omega^n\rangle=2\alpha\int_0^\infty d\omega \; \omega^{n-1}\left[ \left( \frac{1}{2} + \left( \frac{\omega}{\omega_p\gamma} \right)^2 \right) \times \log\left( 1+\left( \frac{\omega_p\gamma}{\omega} \right)^2 \right) - 1\right]
\end{equation}
where $\omega_p$ is the plasma frequency and $\alpha$ is the fine structure constant. A plot of the spectrum of the transition radiation for various values of $\gamma$ is given in figure \ref{fig:spectrum}.

\subsection{Bremsstrahlung}

Low-energy photons are also emitted by accelerated particles. This  radiative process is called bremsstrahlung.

The differential photon number spectrum per unit angle and per unit energy of low energy photons emitted by a charged particle through bremsstrahlung radiation in a collision with a nucleus is given by \cite{jackson}:
\begin{equation}
\frac{d^2N}{d(\hbar\omega)d\Omega} = \frac{z^2\alpha}{4\pi^2\hbar\omega} \left| \vec{\epsilon}^*\cdot\left( \frac{\vec{\beta}^\prime(K)}{1-\vec{n}\cdot\vec{\beta}^\prime(K)} - \frac{\vec{\beta}(K)}{1-\vec{n}\cdot\vec{\beta}(K)} \right) \right|^2
\end{equation}
where $\alpha$ is the fine structure constant, $z$ is the particle electric charge, $\vec{\epsilon}$  is the polarization vector and $\vec{\epsilon}^*$ is its complex conjugate, $\vec{\beta}(K)$ and $\vec{\beta}^\prime(K)$ are the particle speed before and after the emission of the photon and $\vec{n}$ is the unit vector in the direction of the photon. 

The bremsstrahlung cross section is depressed with particle mass $m$ as $m^{-2}$, therefore this process is relevant for electrons only in the energy range of interest for the simulations of the LISA TM charging process. Therefore, in the following we will focus only on electrons emitted or impinging on the LISA TMs.

The discussion on bremsstrahlung is presented for both relativistic and non relativistic electrons.

In the case of non-relativistic or mildly-relativistic secondary electrons produced by primary GCRs, the above expression simplifies to \cite{jackson}:
\begin{equation}
\frac{d^2N}{d(\hbar\omega)d\Omega}=\frac{\alpha}{4\pi^2\hbar\omega} \left| \vec{\epsilon}^*\cdot \vec{\Delta\beta}(K) \right|^2
\end{equation}
where $\vec{\Delta\beta}(K)$ is the difference vector of the velocities of the electron after and before the emission of the photon. By summing over the photon polarizations and integrating over the angles in the opposite direction with respect to the velocity of the electron ($-\pi/2\leq\theta\leq0$, $0\leq\phi\leq2\pi$), it is found:
\begin{equation}\label{eq:brem1}
\frac{dN}{d(\hbar\omega)} =\frac{\alpha}{6\pi}\frac{|\vec{\Delta\beta}|^2}{\hbar\omega}
\end{equation}

\begin{figure}[ht]
\centering
\includegraphics[scale=0.5]{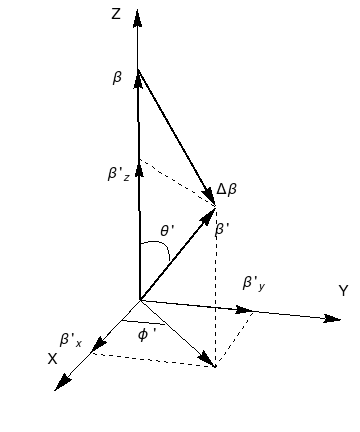}
\caption{Geometry of the calculation of electron velocity variation, $\vec{\Delta\beta}$, before and after bremsstrahlung photoemission.}\label{fig:assi}
\end{figure}

If we assume that the particle initial velocity vector is directed along the z axis, by looking at figure \ref{fig:assi} it can be found that:
\begin{eqnarray}
\vec{\beta}&=(0,0,\beta(K))\\ \nonumber
\vec{\beta}^\prime&=\left( \beta^\prime_x(K),\beta^\prime_y(K),\beta^\prime_z(K) \right)=\\
&=\left( \beta^\prime(K) \cos\phi^\prime \sin\theta^\prime,\beta^\prime(K) \sin\phi^\prime \sin\theta^\prime, \beta^\prime(K) \cos\theta^\prime \right)
\end{eqnarray}
the square of the modulus of the difference vector is given by:
\begin{equation}\label{eq:differenza}
|\vec{\Delta\beta}|^2 = \beta^2(K)+\beta^{\prime2}(K) -2 \beta(K) \beta^{\prime}(K)\cos\theta^\prime
\end{equation}
where $\theta^\prime$ is the angle between the initial and final directions of the electron velocity (see figure \ref{fig:assi}).  $\cos\theta^\prime$ can be calculated by using the conservation of the four momentum: by calling $P_i$  and $P_f$ the initial and final four momenta of the electron, respectively, $P_\gamma$ the four momentum of the emitted photon that is a null vector, and being the electron mass $m$, it follows that $P_i^2=P_f^2=-m^2c^2$ and:
\begin{equation}\label{eq:coseno}
\eqalign{ &P_i-P_f=P_\gamma  \\
 \Rightarrow&P_i^2+P_f^2-2P_i\cdot P_f=P_\gamma^2\\
 \Rightarrow&-2m^2c^2 + 2\frac{E\, E^\prime}{c^2} = 2m^2\vec{v}(K)\cdot\vec{v}^\prime(K)\\
 \Rightarrow&m^2c^4-K\, E^\prime=-m^2c^4\beta(K)\beta^\prime(K) \cos\theta^\prime}
\end{equation}
where the definition of the four momentum $P=(K/c;m\vec{v}(K))=(K/c;\vec{\beta}(K)mc)$ was used and a prime indicates quantities after the emission of a photon with energy $\hbar\omega$.

The modulus  of the speed of the electron of kinetic energy $K$  before  the emission can be calculated from the relation:
\begin{equation}
\gamma(E)m c^2=\frac{m c^2}{\sqrt{1-\beta^2(K)}}=K+m c^2
\end{equation}
where $\gamma(E)$ is the Lorentz factor. By inverting the above expression, one gets:
\begin{equation}\label{eq:inizio}
\beta(K)=\sqrt{\frac{K(2m c^2+K)}{(K+m c^2)^2}}
\end{equation}
and analogously, after the emission of a photon with energy $\hbar \omega$, one finds:
\begin{equation}\label{eq:dopo}
\beta^\prime(K)=\sqrt{\frac{(K-\hbar\omega)(2m c^2+K-\hbar\omega)}{(K-\hbar\omega+m c^2)^2}}.
\end{equation}

\begin{figure}[ht]
\centering
\includegraphics[scale=0.8]{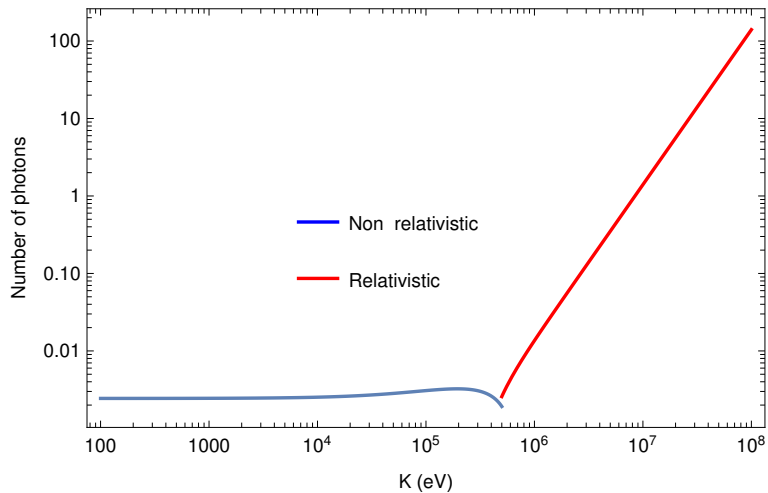}
\caption{Number of photons with energy $4.3\leq \hbar\omega\leq 100$ eV emitted through bremsstrahlung by an electron of energy $K$. The blue curve represents the non relativistic case of equation \eref{eq:brem2} while the red curve represents the relativistic case of equation \eref{eq:brem3}.}\label{fig:brem}
\end{figure}

By substituting $\cos\theta^\prime$ from equation \eref{eq:coseno} and the velocities \eref{eq:inizio} and \eref{eq:dopo} into \eref{eq:differenza} and then into \eref{eq:brem1}, one gets:
\begin{equation}\label{eq:brem2}
\eqalign{
\frac{dN}{d(\hbar\omega)} &= \frac{\alpha}{6\pi}\frac{1}{\hbar\omega} \Bigg( \frac{2mc^2 K+ K^2}{(mc^2+K)^2} + \frac{2mc^2 (K-\hbar\omega)+ (K-\hbar\omega)^2}{(mc^2+K-\hbar\omega)^2} +\\
&+ 2\frac{m^2c^4-K(K-\hbar\omega)}{m^2c^4} \Bigg).}
\end{equation}
Finally, the number $N$ of photons emitted in the range $4.3\leq\hbar\omega\leq100$ eV as a function of the electron energy $K$ can be estimated by integrating over the energy range of the photon given above (blue curve in figure \ref{fig:brem}). The number of photons is also reported in the left side of table \ref{tab:brem} for some values of the energy $K$.

In the case of relativistic electrons, equations \eref{eq:brem1} and \eref{eq:brem2} must be replaced by the following expression \cite{jackson}:
\begin{equation}\label{eq:brem3}
\eqalign{\frac{dN}{d(\hbar\omega)} &= \frac{\alpha}{12\pi}\, \frac{4+\beta(K)+\beta(K)^2}{1+\beta(K)} \frac{|\vec{\Delta\beta}|^2}{\hbar\omega}=\\
 &=\frac{\alpha}{12\pi}\frac{4+\beta(K)+\beta(K)^2}{\hbar\omega(1+\beta(K))}\Bigg( \frac{2mc^2 K+ K^2}{(mc^2+K)^2} + \frac{2mc^2 (K-\hbar\omega)+ (K-\hbar\omega)^2}{(mc^2+K-\hbar\omega)^2} \\
&+ 2\frac{m^2c^4-K(K-\hbar\omega)}{m^2c^4} \Bigg) }
\end{equation}
where $\beta(K)$ is given by equation \eref{eq:inizio}. Again, this expression must be integrated over the energy of the photon between 4.3 eV and 100 eV. The number $N$ of photons emitted by a relativistic electron in the considered range of energies is represented by the red curve of figure \ref{fig:brem}. On the right side of table \ref{tab:brem} the number of photons  calculated with equation \eref{eq:brem3} at some values of the energy $K$ is reported.

\begin{table}[ht]
\centering
\begin{tabular}{lc|lc}
\multicolumn{2}{c|}{Non relativistic electrons}&\multicolumn{2}{c}{Relativistic electrons}\\
\hline
Energy (eV) & N & Energy (eV) & N\\
\hline
$10^3$         & $2.45\times 10^{-3}$ & $10^6$         & 0.01 \\
$5\times 10^3$ & $2.48\times 10^{-3}$ & $5\times 10^6$ & 0.35 \\
$10^4$         & $2.53\times 10^{-3}$ & $10^7$         & $1.40$ \\
$5\times 10^4$ & $2.83\times 10^{-3}$ & $5\times 10^7$ & 35.0 \\
$10^5$         & $3.08\times 10^{-3}$ & $10^8$         & 140
\end{tabular}
\caption{Number of photons with energies varying between 4.3 eV and 100 eV produced by electrons through bremsstrahlung. The numbers on the left are calculated using the non relativistic equation \eref{eq:brem2}, the numbers on the right are calculated using the relativistic equation \eref{eq:brem3}.}\label{tab:brem}
\end{table}

\begin{figure}[ht]
\centering
\subfigure{\includegraphics[scale=0.6]{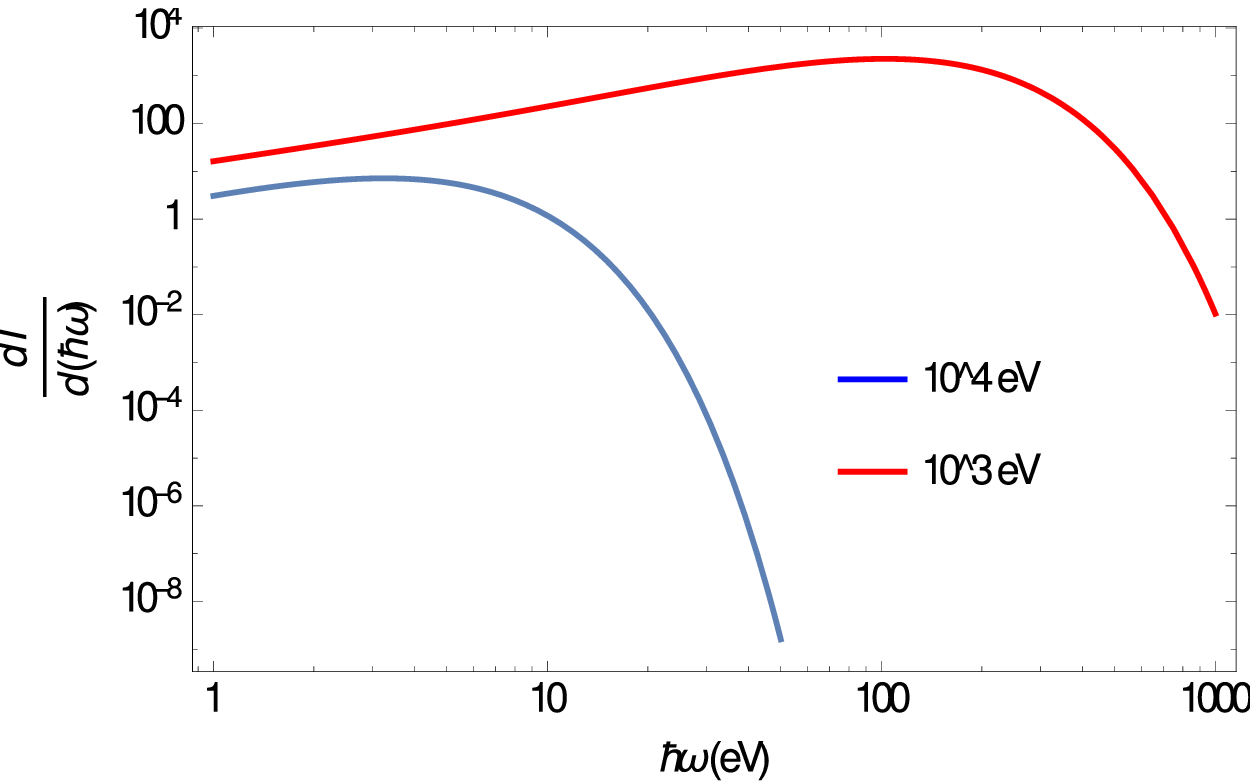}}\subfigure{\includegraphics[scale=0.6]{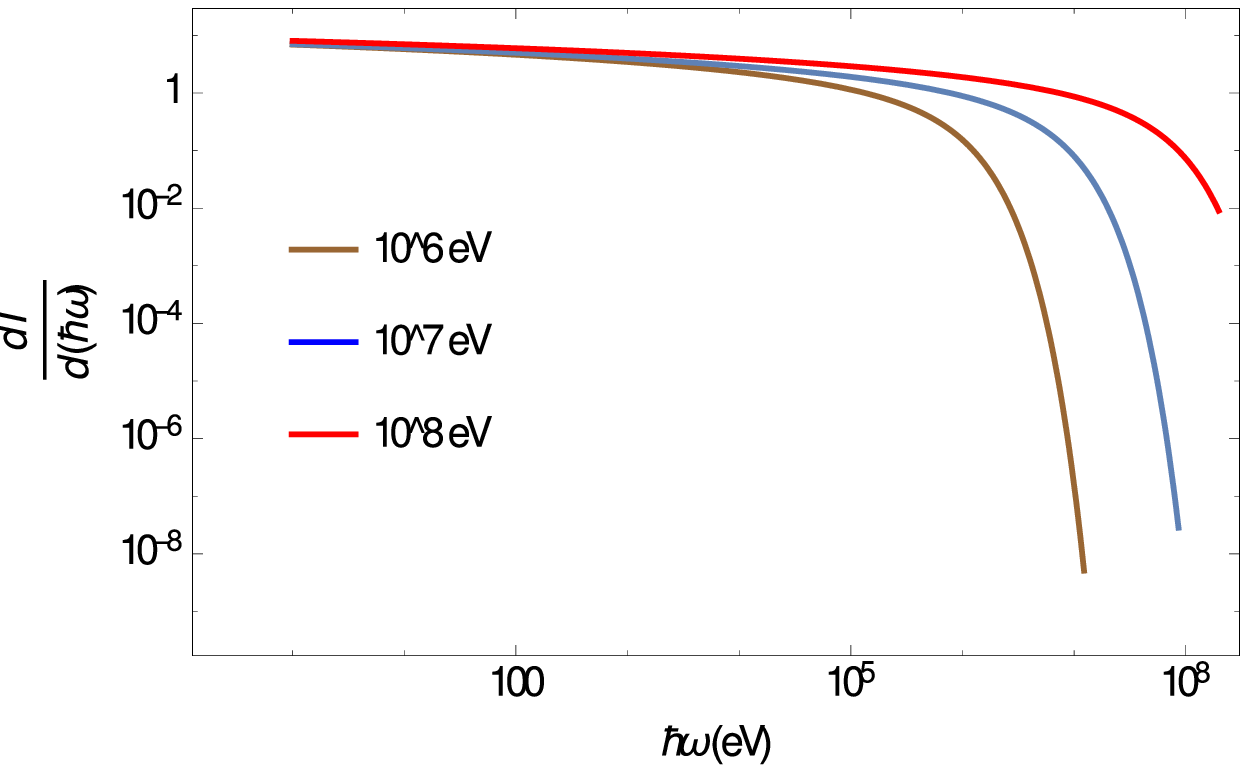}}
\caption{Spectrum of bremsstrahlung radiation emitted in a single collision by a non relativistic electron (left) and a relativistic one (right) for various electron energies.}\label{fig:brem_sp}
\end{figure}

The energy spectrum of bremsstrahlung radiation for a non relativistic electron of velocity $v$ that hits an atom of mass $M$ and atomic number $Z$ is given by \cite{landau_mech}:
\begin{equation}\label{eq:sp_nrel}
\frac{dI}{d(\hbar\omega)}=6.2 \times 10^{11}\;\frac{\pi Z e^4 \mu^2  \omega^3}{6c^3\hbar^4 K^2} \left( \frac{1}{m_e} - \frac{Z}{M} \right)^2 \left\{ \left[ H^{(1)\prime}_{i \nu}(i\nu\epsilon) \right]^2 - \frac{\epsilon^2-1}{\epsilon^2} \left[ H^{(1)}_{i \nu}(i\nu\epsilon) \right]^2 \right\}
\end{equation}
where $K=m_e v^2/2$ is the electron energy, $\mu$ is the reduced mass of the system electron-atom, $m_e$ and $e$ are the electron mass and charge, $H^{(1)}_a(x)$ is the Hankel function of order $a$ and where:
\begin{equation}
\nu=\frac{\omega e^2Z}{\hbar \mu v^2}
\end{equation}
and 
\begin{equation}
\epsilon=\sqrt{1+\frac{2K\mathcal{M}}{\mu e^2Z}}
\end{equation}
where $\mathcal{M}$ is the angular momentum of the system electron-atom. The factor $6.2 \times 10^{11}$ is the conversion coefficient between erg and eV. A log-log plot of equation \eref{eq:sp_nrel} is reported on the left side of figure \ref{fig:brem_sp}.

In the relativistic case, the energy spectrum is given \cite{jackson}:
\begin{equation}\label{eq:sp_rel}
\frac{dI}{d(\hbar\omega)}=6.2 \times 10^{11}\;\frac{2}{\pi} \frac{e^2}{\hbar c} \frac{1}{\beta^2} \left[ x K_0(x)K_1(x) - \frac{\beta^2}{2} x^2 \left( K_1(x)^2-K_0(x)^2 \right) \right]
\end{equation}
where $K_n$ are the modified Bessel  functions of order $n$,
\begin{equation}
x=\frac{\omega b_{min}}{c \hbar\gamma \beta}
\end{equation}
and
\begin{equation}
b_{min}=\frac{\hbar}{c\gamma m_e \beta}.
\end{equation}
A log-log plot of equation \eref{eq:sp_rel} is reported on the right side of figure \ref{fig:brem_sp}.

{The main contributions to photon production is given by relativistic electrons  producing photons below 100 eV. This process will be taken into account in the future simulations.}

\subsection{Photoelectric effect}
At low energies photon electron production occurs through photoelectric effect only. Photoelectric effect is the emission of an electron from a material due to the absorption of a photon. This process occurs if the photon has an energy larger than the work function of the material in order to extract an electron; for gold this threshold is 4.3 eV, as it was recalled above. This process has a quantum efficiency lower than 1. In particular, in the energy range from 4.3 eV to 100 eV, the quantum efficiency varies from 3\% to 10\% \cite{quant_eff,quant_eff2}. As a result, the estimated contribution of low-energy photons to photoelectron production appears negligible below 100 eV and therefore can be neglected in the future TM charging simulations for the LISA interferometer.

\section{Quantum Mechanics processes and LISA TM charging}
\label{sec:QMeff}

QM effects may take place in the interactions between electrons in the tens of eV energy range and the surfaces of the gold-coated TMs and electrodes. These effects interfere with the charging process through electron backscattering and  diffraction. To this purpose, possible recrystallization of the TM after baking may affect the number of electrons deposited on the TMs by changing the characteristics of the crystal lattice of the surface of the material.

\subsection{Low-energy electron diffraction}
\label{sec:diff}
The most important QM effect to be considered is the diffraction of low-energy electrons by the crystal lattice of the TMs. This effect was first studied by Davisson and Germer in 1927 \cite{DG}. They used an electron gun to accelerate electrons with a potential varying between 25 and 75 V. Electrons impinged perpendicularly on a block of nickel. Initially Davisson and Germer did not observe any diffraction pattern: this was because the nickel block was exposed to air in a previous experiment and oxidized. After the block was baked ``\emph{at various high temperatures in hydrogen and in vacuum}'' \cite{DG}, they could see a diffraction pattern. This happened because baking caused a recrystallization of the nickel block: before this process, the electrons were impinging on a large number of small crystals, while after baking the impact occurred on a smaller number of large crystals and this caused the diffraction to take place. The diffraction pattern was evident when the energy of the electrons was varying  between 40 and 68 eV, with an enhancement of the process at 54 eV, see figure 10 in \cite{DG}, i.e. when the de Broglie wavelength of the electrons and the lattice spacing of the metal were comparable.

\begin{figure}[h]
\centering
\includegraphics[scale=0.5]{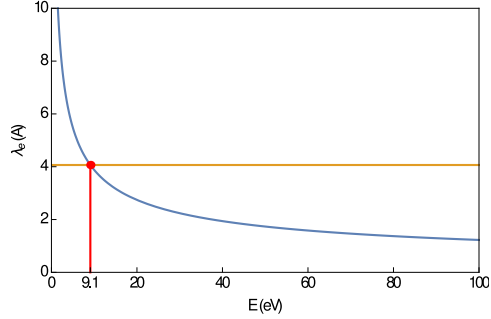}
\caption{De Broglie wavelength $\lambda_e$ of an electron as a function of the electron energy (blue line). The horizontal brown line represents the nominal spacing of the gold-platinum lattice ($\approx4$ {\AA}). The red line marks the energy at which de Broglie wavelength and the lattice spacing are equal ($\approx 9.1$ eV).}\label{fig:spacing}
\end{figure}

For the gold-coated TMs of LPF and LISA, the lattice spacing at the surface is about 4 {\AA} \cite{gold}; the de Broglie wavelength of an electron has approximately the same value at about 9.1 eV {as it can be noticed in figure \ref{fig:spacing}}. There is, however, an interval of energy  $\Delta K$ around 9.1 eV for which diffraction occurs. This interval of energies is calculated according to the imaginary part of the atomic potential, as illustrated in the following \cite{libro2}. When the electron enters the material, the momentum $k$ is given by:
\begin{equation}
\frac{\hbar^2k^2}{2m}=K+U_{0r}+i U_{0i}
\end{equation}
where $U_{0r}+i U_{0i}$ is the complex potential energy inside the material and $m$ the electron mass. From the above expression, one can see that the momentum must be a complex number with real and imaginary parts given respectively by:
\begin{equation}\label{eq:prima}
k_r=\sqrt{\frac{2m}{\hbar^2}(K+U_{0r})}
\end{equation}
and
\begin{equation}\label{eq:seconda}
k_i= \frac{m}{\hbar^2} \frac{U_{0i}}{k_r}.
\end{equation}
By substituting \eref{eq:prima} into \eref{eq:seconda} and recalling that $k_i=\lambda^{-1}(K)$, where $\lambda(K)$ is the inelastic mean free path of the electron in the material \cite{libro2}, the imaginary part of the inner potential is determined as follows:
\begin{equation}
U_{0i}=\lambda^{-1}(K)\,\sqrt{\frac{2\hbar^2}{m}(K+U_{0r})}.
\end{equation}
$U_{0i}$ allows to calculate the spread in energy corresponding to the diffraction peak since $\Delta K=U_{0i}$ \cite{libro2}. Using the value of $U_{0r}=15$ eV given in \cite{V0i1,V0i2} and the value of the mean free path at 9.1 eV from \cite{libro2} ($\lambda\approx 20$ {\AA}),  $\Delta E\approx 1.74$ eV, therefore diffraction occurs approximately between 7 and 12 eV. A lower limit of 12 eV for electron propagation in the future LISA Monte Carlo simulations will be consequently set.

\subsection{Low-energy electron backscattering}

Backscattering of electrons is another QM effect to take into proper consideration since not all the electrons emitted from the electrodes will end up on the TMs. To calculate the scattering amplitude, we follow the discussion reported in \cite{libro2} and assume that the electrons propagate as spherical waves. After scattering, the asymptotic wavefunction is given by:
\begin{equation}\label{eq:scattering}
\exp(-i\vec{k}\cdot\vec{r}) + f(\theta) \frac{\exp(-i\vec{k}\cdot\vec{r})}{r}
\end{equation}
where $f(\theta)$ is the scattering amplitude given by:
\begin{equation}\label{eq:amplitude}
f(\theta)=-4\pi \sum_{l=0}^{l_{max}} (2l+1) t_l P_l(cos(\theta))
\end{equation}
where $P_l(x)$ are the Legendre polynomials and:
\begin{equation}\label{eq:tl}
t_l=\frac{\hbar^2}{2m} \frac{1}{2ik_r} \left[ \exp(2i\delta_l)-1 \right]
\end{equation}
where $\exp(2i\delta_l)$ is the phase shift and $k_r$ is given by \eref{eq:prima}. The maximum angular momentum $l_{max}$ to be included in the calculation of the scattering amplitude can be estimated by using the Babinet principle, i.e. by dividing the crystal lattice spacing $d$ by the de Broglie wavelength $\lambda_e$ of an electron with energy $K$ \cite{libro2}:
\begin{equation}
l_{max}\approx \pi \frac{d}{\lambda_e(K)}.
\end{equation}

\begin{figure}
\centering
\subfigure{\includegraphics[scale=0.5]{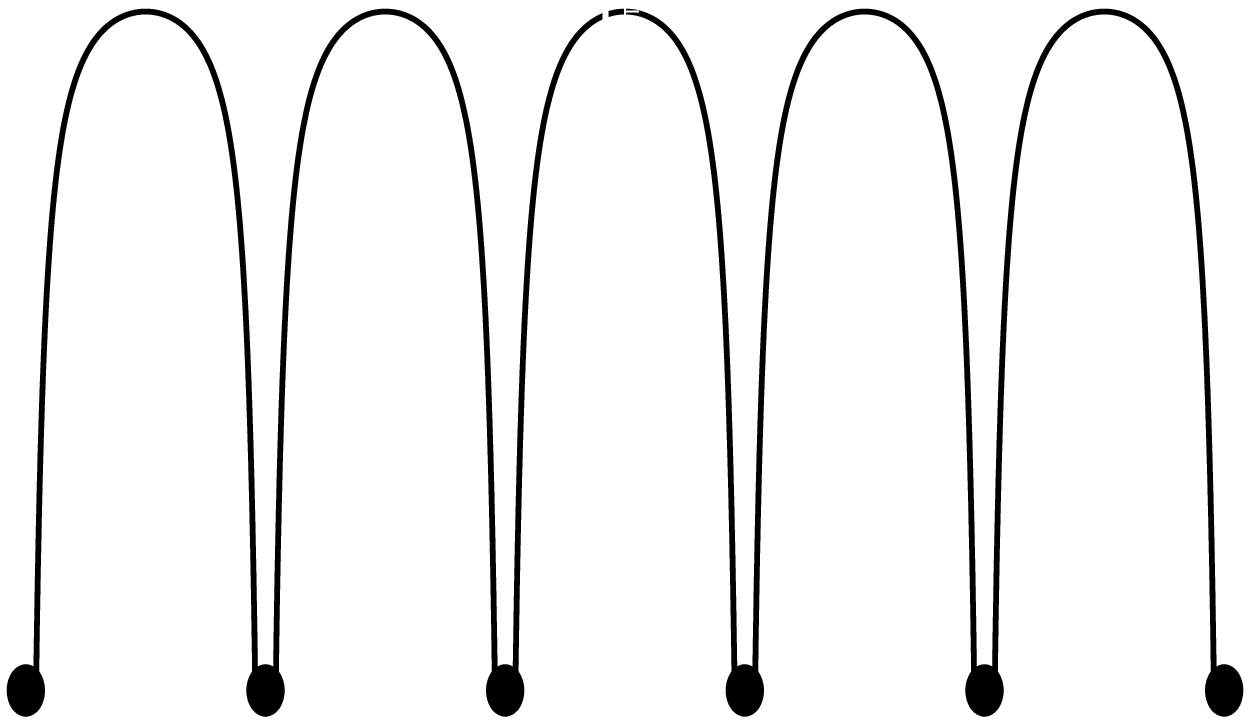}}\subfigure{\includegraphics[scale=0.5]{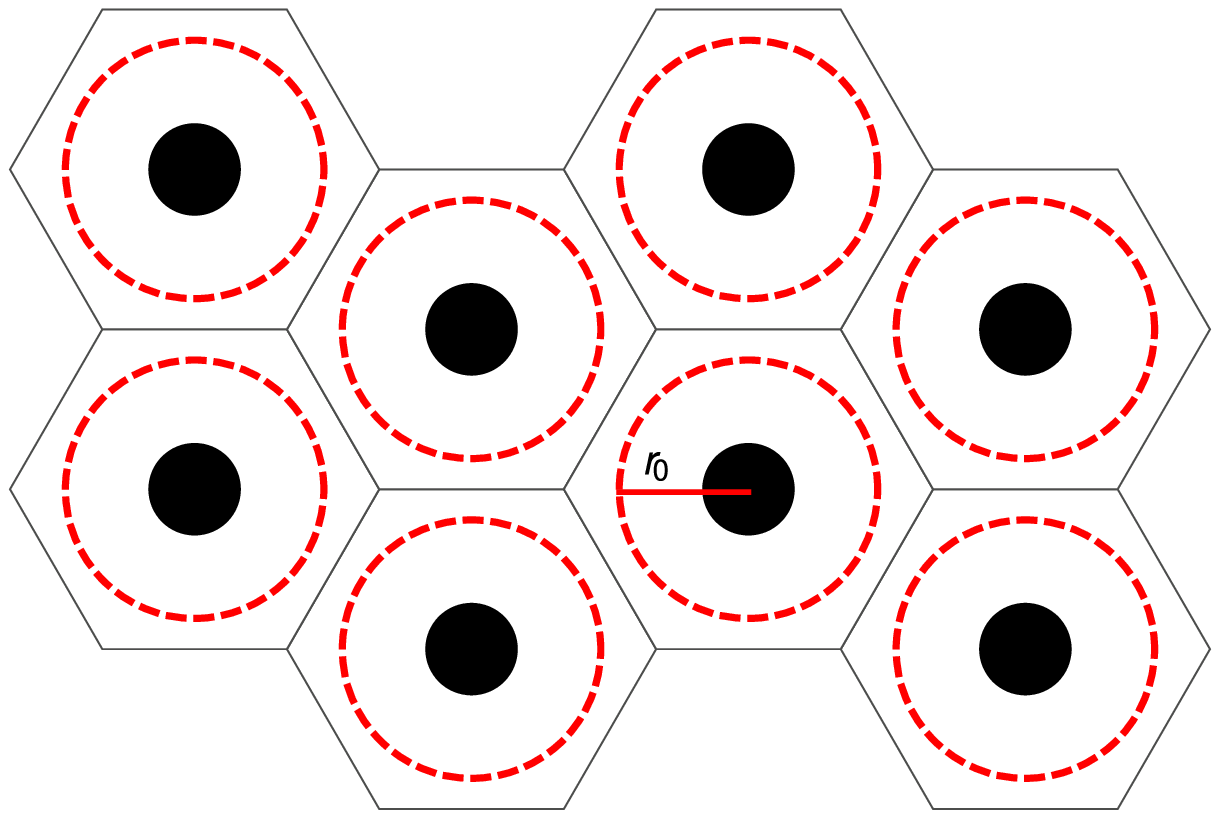}}
\caption{Schematic representation of the \emph{muffin-tin} potential. The left image represents the potential in the lattice: the potential well corresponds to the position of the ions represented by black disks; the right image shows the position of the ions in the lattice. The red dashed circles represent the domain for the solution of the radial Schr\"odinger equation \eref{eq:shcr}; also in red is indicated the radius $r_0$ of one of such spheres. See also \cite{libro2,ashcroft}.}\label{fig:muffin_tin}
\end{figure}

In order to calculate the phase shift, the Schr\"odinger equation of an electron propagating inside the crystal potential described by the muffin-tin model must be solved (a schematic representation of the muffin-tin potential is given in figure \ref{fig:muffin_tin}; see also \cite{ashcroft,libro2}). The Schr\"odinger equation was numerically solved into the potential well of one atom by assuming that the potential has spherical symmetry and that the spheres around each ion do not overlap, i.e.  the radius of each sphere $r_0$  must be smaller than half of the lattice spacing (see the right part of figure \ref{fig:muffin_tin}). One must then  match the resulting radial eigenfunction $R_l(r)$ with the eigenfunction $S_l(r)$ of the electron propagating in the space between two atoms where the potential is $V_{0r}$. 

With the assumption of spherical symmetry, the radial and angular part of the Schr\"odinger equation for an electron interacting with the potential of one gold atom can be separated; the angular part can be found for example in \cite{messiah} and its eigenfunctions are spherical harmonics, while the radial equation is reported in \cite{libro2}:
\begin{equation}\label{eq:shcr}
\eqalign{
&-\frac{\hbar^2}{2m} \frac{1}{r^2} \frac{d}{dr}\left( r^2 \frac{d}{dr} \right) R_l(r) + \frac{\hbar^2}{2m} \frac{l(l+1)}{r^2} R_l(r) +\\
&- \frac{Ze^2}{r} R_l(r) + V_{ex}(r) R_l(r)=K R_l(r)}
\end{equation}
where $Z$ is the material atomic number, $l$ is the eigenvalue of the angular momentum, $m$, $e$ and $K$ are the electron mass, charge and energy, respectively, and  $V_{ex}(r)$ is the exchange potential i.e. the contribution to the potential given by the electron-electron interaction that takes into account the Pauli exclusion principle and can be approximated by \cite{libro2,exch1,exch2}:
\begin{equation}
V_{ex}(r)=-2\left( \frac{3}{8\pi} \rho(r) \right)^{1/3}
\end{equation}
where $\rho(r)$ is the electron density of the gold atom which, for material with high $Z$ like gold, can be calculated with the Thomas-Fermi theory \cite{landau}:
\begin{equation}
\rho(r)= Z^2 \frac{32}{9\pi^3} \left[ \chi\left(\frac{r}{a_b}\frac{Z^{1/3}}{0.885}\right) \frac{a_b 0.885}{r Z^{1/3}} \right]^{3/2}
\end{equation}
where $a_b$ is the Bohr radius and where the function $\chi(x)$ is given by the differential equation:
\begin{equation}
\frac{d^2\chi(x)}{dx^2}=\frac{\chi(x)^{3/2}}{\sqrt{x}}
\end{equation}
with the boundary conditions: $\chi(0)=1$ and $\chi(\infty)=0$. Equation \eref{eq:shcr} must be solved numerically for each value of $l<l_{max}$ with the condition that $R_l(r)$ is bounded at the origin.

Under the assumption that the electron propagates as a spherical wave, the asymptotic eigenfunction of the electron propagating in the space between two atoms where the potential is $V_{0r}$, is given by a linear combination of Hankel functions $h_l^{(1)}$ and $h_l^{(2)}$ \cite{libro2}:
\begin{equation}\label{eq:hank}
S_l(r)= \frac{1}{2} \left[ \exp(2i\delta_l) h_l^{(1)}(k_rr)+h_l^{(2)}(k_rr) \right].
\end{equation}

\begin{figure}[htb]
\centering
\subfigure{\includegraphics[scale=0.65]{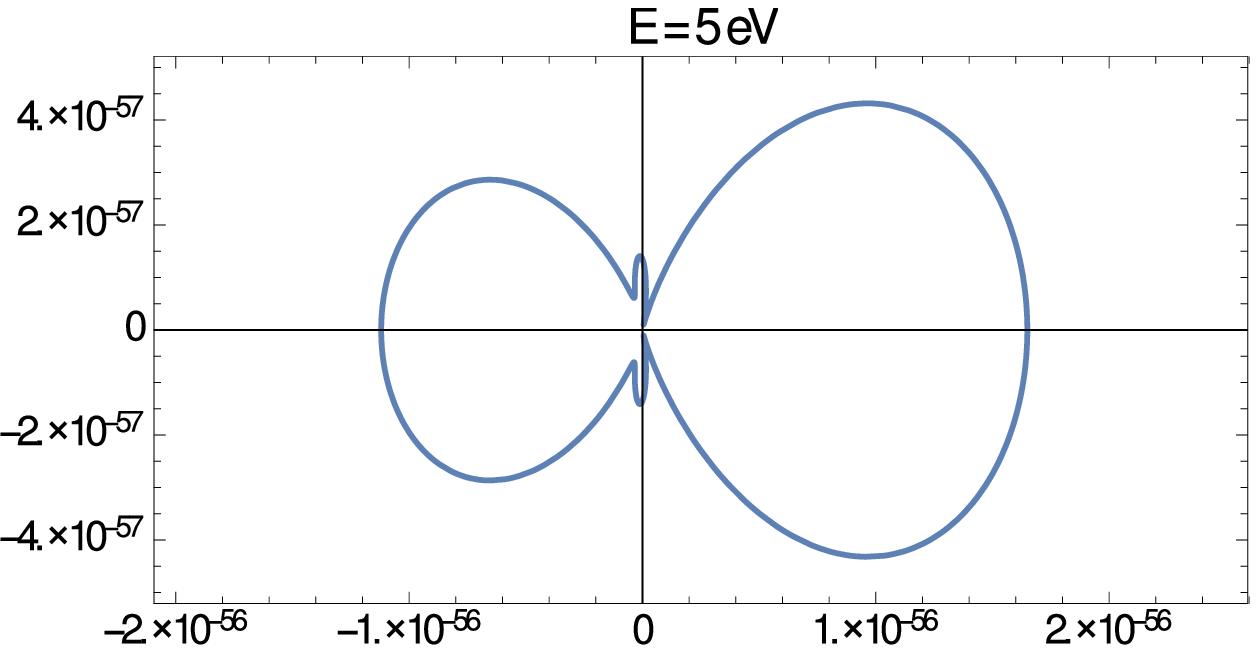}}\subfigure{\includegraphics[scale=0.68]{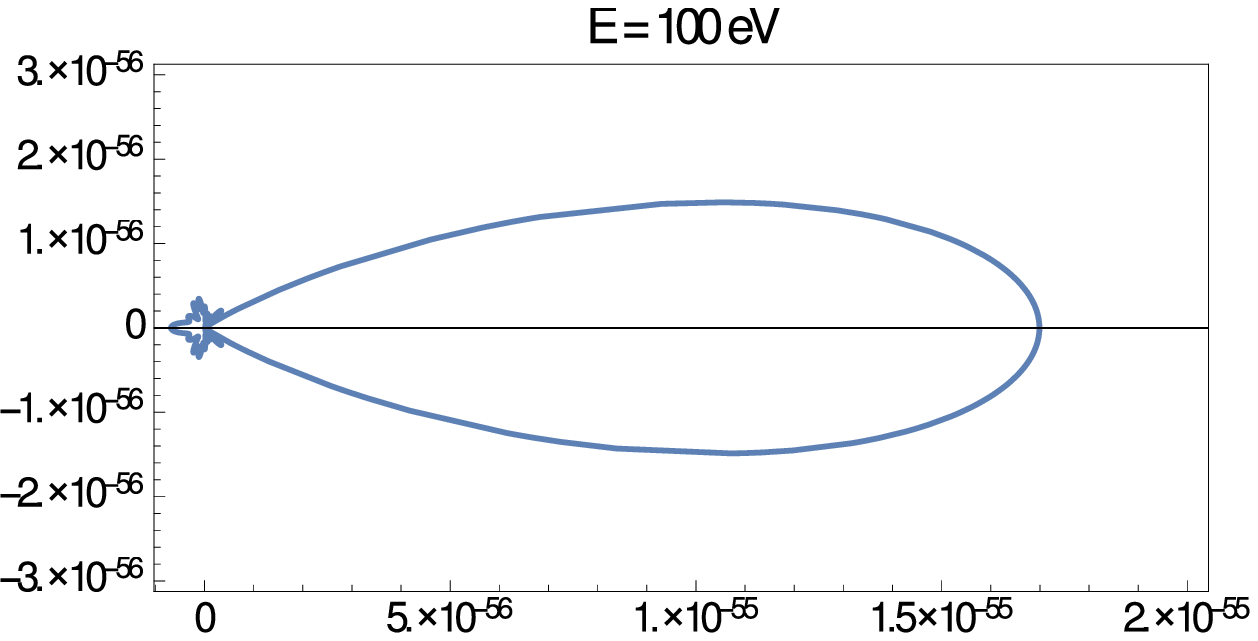}}
\caption{Polar plots of $|f(\theta)|^2$ for electron energies of 5 eV (left) and 100 eV (right).}\label{fig:f}
\end{figure}

By matching the logarithmic derivatives of the two functions $R_l(r)$ and $S_l(r)$ evaluated at the radius $r_0$ of the sphere that contains the ion, i.e. by calculating:
\begin{equation}
\frac{R^\prime_l(r_0)}{R_l(r_0)}=\frac{S^\prime_l(r_0)}{S_l(r_0)} = \frac{\exp(2i\delta_l)h_l^{\prime(1)}(k_rr_0) +h_l^{\prime(2)}(k_rr_0)}{\exp(2i\delta_l)h_l^{(1)}(k_rr_0) +h_l^{(2)}(k_rr_0)}
\end{equation}
where a prime indicates  deriving formulas for the radial coordinate, the phase $\exp(2i\delta_l)$ appearing in \eref{eq:tl} is calculated and then the scattering amplitude $f(\theta)$ and its square $|f(\theta)|^2$ are estimated. This last term is strongly dependent on the electron energy as it can be seen from figure \ref{fig:f} for electron energies of 5 eV and 100 eV. In order to find the probability that an incoming electron is backscattered, one must integrate  $|f(\theta)|^2$ between $\pi/2$ and $3/2\pi$; the backscattering probability thus calculated is reported in table \ref{tab:backscattering} as a function of the electron energy from 5 eV to 100 eV.

{Despite this process appears relevant, electron backscattering from TMs and electrodes may compensate in the TM charging. The possible differences between simulations outcomes and LPF TM charging measurements will provide precious clues about the role of QM effects. }

\begin{table}[ht]
\centering
\begin{tabular}{cc|cc}
Energy (eV) & \% backs. &Energy (eV) & \% backs.\\
\hline
5 & 41.8 & 55 & 15.3 \\
10& 28.2 & 60 & 17.8 \\
15& 26.1 & 65 & 16.3 \\
20& 22.7 & 70 & 14.8 \\
25& 22.8 & 75 & 16.3 \\
30& 20.0 & 80 & 16.1 \\
35& 18.3 & 85 & 15.3 \\
40& 17.6 & 90 & 14.2 \\
45& 18.0 & 95 & 13.3 \\
50& 16.4 & 100& 13.5 \\
\end{tabular}
\caption{Percentage of backscattered electrons as a function of the electron energy.}\label{tab:backscattering}
\end{table}

\subsection{Recrystallization}
As it was discussed in section \ref{sec:diff}, the baking of the nickel block in the Davisson and Germer experiment was a crucial step in the discovery of electron diffraction; the importance of the baking is ascribable to the changes in the crystal lattice and in the material surface that occur at high temperature, this means that material lattice and surface characteristics play a crucial role in the quantum properties of the surface of the material.

For a gold test mass pure at 99.9999\%, the recrystallization takes place between 150 C and 200 C \cite{gold2}; in the case of LPF, the gold-coated TMs were baked at 120 C for a day and therefore, nominally, no effects on the crystal lattice and on the surface are expected.

\section{Summary of low-energy electromagnetic processes affecting the LISA test mass charging}
\label{sec:summary}
{The LPF TM effective charging measured in spring 2016 was of 1000-1300 e s$^{-1}$. In table \ref{tab:ris}, the relative contribution to the TM charging of the low-energy electromagnetic processes described in this work are summarized. These estimates are meant for incident GCR protons constituting 90\% of the cosmic-ray bulk. In case of SEP occurrence the values reported in table \ref{tab:ris} would increase proportionally to the event fluence. The outcomes of this work, despite being obtained for the TM charging of the LISA-like space interferometers may result of interest by other communities for dose estimates absorbed in space and for applied surface physics problems since the effect of particles with energies ranging between a few tens of MeV and thousands of GeV cannot be easily studied on single beam tests.}

\begin{table}[ht]
\centering
\begin{tabular}{cc}
Process & Number of charges per second \\
\hline
Ionization & Several Hundreds\\
IIEE & Tens \\
EIEE & Hundreds \\
Sputtering & A few \\
Transition radiation & A few \\
Bremsstrahlung & A few
\end{tabular}
\caption{Relative role of the low-energy electromagnetic  processes affecting the LISA TM charging in terms of number of rate of charges deposited and escaping the TMs.}\label{tab:ris}
\end{table}

\section{Conclusions}
\label{sec:conc}
The LPF TM charging simulations carried out before the mission launch indicated a net charging of the TM in very good agreement with observations, while the shot noise was underestimated by a factor 3-4. In this paper, transport of electrons and photons $<$100 eV, neglected in the LPF Monte Carlo Geant4 and FLUKA  simulations, have been discussed for LISA and future space interferometers. Ionization, kinetic electron emission, atomic sputtering, bremsstrahlung and photoelectric effect due to low-energy photons and  quantum mechanics effects have been discussed in detail.

{Among low-energy electromagnetic processes ($<$ 100 eV), the most relevant role appears to be played by particle ionization that increases the charge deposited and escaping the TMs by several hundreds in addition to the few hundreds obtained with FLUKA and Geant4 simulations by considering the incident GCR protons above 100 MeV. In comparison to ionization, IIEE and EIEE increase the effective charging by hundreds of charges per second; the other processes give a more modest contributions of a few tens  charges deposited per second}. Low-energy photons  may produce low-energy electrons through photoelectric effect, however even by including the effects of transition radiation  and bremsstrahlung from the overall GCR sample, the contribution of these processes to the TM effective charging is at most of a few electrons per second and, consequently, can be disregarded.  QM effects do not lead to electron production, but may affect the number of charges that end up on the TMs because of the backscattering of  electrons especially at very low energies. {However, backscattering from electrodes and TMs may compensate}. The role of QM effects on TM charging may be investigated by considering the same physical processes  described for LISA in the simulations of LPF for measurement comparison.  The role of baking of the TMs up to 120 C is supposed to be negligible from the point of view of QM effects.

New dedicated Monte Carlo simulations will be carried out with FLUKA and Geant4 in addition to the LEI Monte Carlo for LISA.

\end{document}